\documentclass[aps,prl,twocolumn,groupedaddress]{revtex4}

\usepackage{graphicx}

\begin{document}


\title{Ultra-low noise microwave generation with fiber-based optical frequency comb and application to atomic fountain clock}


\author{J. Millo, M. Abgrall, M. Lours,  E.M.L. English, H. Jiang, J. Gu{\'e}na, A. Clairon, S. Bize, Y. Le Coq}
\email{yann.lecoq@obspm.fr}
\author{G. Santarelli}
\affiliation{LNE-SYRTE, Observatoire de Paris, CNRS, UPMC, 61 Avenue de
l'Observatoire, 75014 Paris, France}
\author{M.E. Tobar}
\affiliation{School of Physics, University of Western Australia, Crawley 6009, Australia}


\date{\today}

\begin{abstract}

We demonstrate the use of a fiber-based femtosecond laser locked onto an ultra-stable optical cavity to generate a low-noise microwave reference signal. Comparison with both a liquid Helium cryogenic sapphire oscillator (CSO) and a Ti:Sapphire-based optical frequency comb system exhibit a stability about $3\times10^{-15}$ between 1\,s and 10\,s. The microwave signal from the fiber system is used to perform Ramsey spectroscopy in a state-of-the-art Cesium fountain clock. The resulting clock system is compared to the CSO and exhibits a stability of $3.5\times10^{-14}\tau^{-1/2}$. Our continuously operated fiber-based system therefore demonstrates its potential to replace the CSO for atomic clocks with high stability in both the optical and microwave domain, most particularly for operational primary frequency standards.

\end{abstract}

\pacs{}

\maketitle


Atomic fountain frequency standards based on cold atoms are the most widely used high accuracy atomic clocks \cite{WynandsMetrologia2005}. About ten fountains currently participate to the definition of the SI second at a level of $10^{-15}$ or better \cite{FountainItalians, FountainGB, FountainUS, FountainSYRTE}. Besides classical metrological and timekeeping tasks, accurate and stable atomic fountain clocks can also perform high precision fundamental physics tests \cite{OldAlphaDot, AlphaDot, LorentzInvariance}.

State-of-the-art microwave atomic fountain clocks \cite{FountainSYRTE} exhibit quantum projection noise (QPN) \cite{Santarelli1999} limited short term stability well below $10^{-13}$ at 1\,s integration time. However, the intrinsic phase noise of the microwave signal used as interrogation oscillator for these atomic standards degrades performances from the fundamental QPN limit, via the Dick effect \cite{Dick, Santarelli1998}. Therefore, the realisation of extremely low noise microwave oscillators is of prime importance for frequency standards to reach high stability \cite{Santarelli1999}. Other applications of low noise microwave sources include radar, deep space navigation \cite{DeepSpace} and ultra-high resolution very-long-baseline interferometry (VLBI) \cite{VLBI}.

The interrogation oscillator for the LNE-SYRTE atomic fountain clocks is currently a liquid Helium Cryogenic Sapphire Oscillator (CSO) at 11.932\,GHz \cite{Tobar}. A CSO was until recently the only available technology allowing QPN limited stability of fountain clocks at a few $10^{-14}$ at 1\,s. The cost of operation and maintenance associated with cryogenic cooling make it desirable to find an alternative technique. Optical ultra-stable reference cavities \cite{CavityYe2007, CavityWebster2008}, on the other hand, offer reliable and low maintenance high purity source although in the optical frequency range. Transfer of the stability of such optical reference (typically in the lower $10^{-15}$ range at 1\,s) to the microwave domain by use of a Ti:Sapphire based optical frequency comb (TSOFC) has been demonstrated with a residual instability of $6.5\times10^{-16}$ at 1s \cite{Bartels2005}. However, for long term and reliable operation, fiber-based optical frequency combs (FOFC) are more desirable. Lipphardt et al. recently demonstrated microwave generation at the level of instability of $1.2\times10^{-14}$ at 1\,s \cite{PTB_Low_Noise} with such a system. 

Here we present a different technique to generate a low noise microwave from a FOFC and demonstrate an instability in the low $10^{-15}$ range at 1\,s by comparisons with both a CSO and a TSOFC generated microwave. The low noise microwave signal at 11.932\,GHz from the FOFC was used as a replacement of the CSO signal in our frequency synthesis system \cite{Chambon2005} to perform Ramsey spectroscopy in a Cesium fountain and was locked to the atomic signal. The resulting clock was measured against the CSO and demonstrated an instability of $3.5\times10^{-14}\tau^{-1/2}$ for measurement time $\tau$, identical to the one obtained under identical operating conditions using the CSO as fountain interrogation oscillator.

\begin{figure}[ht]
	\includegraphics[width = 0.45 \textwidth]{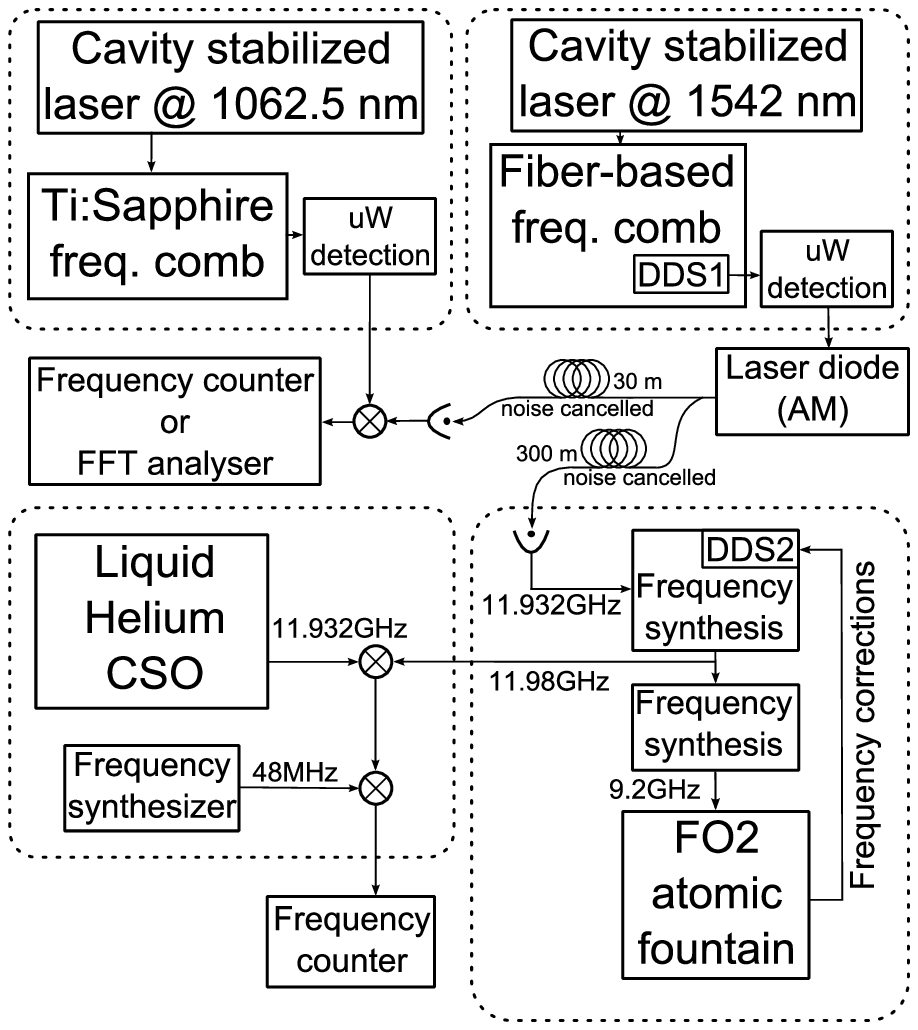}
	\caption{\label{FigSchematic} Schematic of the experimental system. CSO: liquid Helium cooled Cryogenic Sapphire Oscillator. DDS: Direct Digital Synthesizer. FFT: Fast Fourier Transformer analyzer. uW: microwave signal.}
\end{figure}

Essential to this work was our designing and implementing very low vibration sensitivity optical cavities \cite{Millo2008Cavity}. The measured frequency instability of the beat-note signal between two CW fiber lasers at 1542\,nm stabilized on two independent ultra-stable cavities is below $2\times10^{-15}$ at 1\,s \cite{Jiang2008}. A commercial Erbium doped fiber femtosecond laser \footnote{Menlo Systems GmbH, M-Comb + P250 + XPS1500} of repetition rate $f_{\rm rep}\simeq250$\,MHz, with inbuilt f-2f interferometer for measuring the carrier-envelop offset frequency $f_0$ \cite{Newbury_f_2f}, is locked onto one of these reference lasers. The lock technique is as follows: a 30\,mW output port (100\,nm spectral bandwith) from the oscillator is sent through a fibered Bragg grating at 1542 nm whose reflected light (1 nm spectral bandwith) is directed through a circulator to a fibered 50/50 power combiner where it is mixed with the reference ultra-stable reference light of optical frequency $\nu_{\rm cw}$. The resulting beatnote signal $f_{\rm b}=\nu_{\rm cw}-N\times f_{\rm rep}-f_0$ (with N a large integer) is detected on a photodiode and mixed with $f_0$. After filtering, the relevant sideband produces a frequency $\nu_{\rm cw}-N\times f_{\rm rep}$ independent of $f_0$. This signal is cleaned by a tracking oscillator filter (2 MHz bandwidth), divided by 128 and mixed with a reference frequency synthesized by a Direct Digital Synthesizer (DDS) referred as DDS1 (see fig \ref{FigSchematic}) to produce a phase error signal. This signal acts on the pump-power controller of the femtosecond laser through an optimized phase-lock loop filter. The servo bandwith is 120\,kHz, which, combined with the division factor of 128, allows robust and reliable servo-locking. The high gain of the loop between 1\,Hz and 1\,kHz allows the noise to be limited, in principle, to that of the reference CW laser. Once locked onto the optical reference $\nu_{\rm cw}$, the repetition rate $f_{\rm rep}$ (and all its harmonics) reproduces the ultra-high stability of the optical reference transferred in the microwave domain. Our optical reference laser exhibits a long-term drift of a few $10^{-15}\,s^{-1}$ which is removed by a constant feed-forward linear ramp on DDS1.

\begin{figure}[ht]
	\includegraphics[width = 0.45 \textwidth]{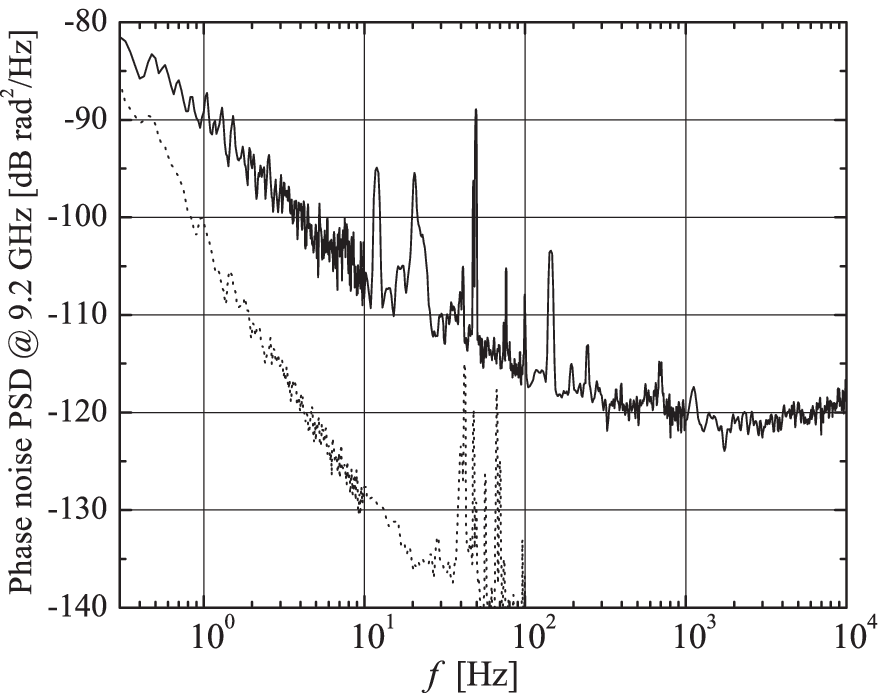}
	\caption{\label{FigPhaseNoise} Phase noise power spectral density (PSD) at 9.2\,GHz of the beatnote between the microwaves generated by the fiber-based and the Ti:Sapphire-based systems. Dashed line: phase noise PSD of two ultra-stable lasers locked onto 1542\,nm ({\em i.e.} 194\,THz) independent reference cavities (scaled to 9.2\,GHz).}
\end{figure}

To generate microwave signals, the transmitted output of the Bragg grating (9\,mW), containing all the spectrum (not useful for generating the beatnote signal $f_{\rm b}$) is sent to a fast InGaAs pigtailed photodiode (Discovery model DSC40S). The output signal of the detector contains all the harmonics of the repetition rate, up to 20\,GHz. In order to characterize and use the microwave from this FOFC in distant laboratories, the harmonic of interest is then filtered, amplified and used to amplitude modulate a pigtailed C-band Telecom diode laser. This amplitude modulated optical signal is transmitted through optical fibre to distant laboratories. To cancel the residual noise of the fiber transmission line, we implemented a round-trip active compensation scheme which is a simplified version of the one described in \cite{OpticalLink}. The performances of this optical link was measured to add an instability well below $3\times10^{-15}\tau^{-1}$ for a few hundred meters optical path.

\begin{figure}[ht]
	\includegraphics[width = 0.45 \textwidth]{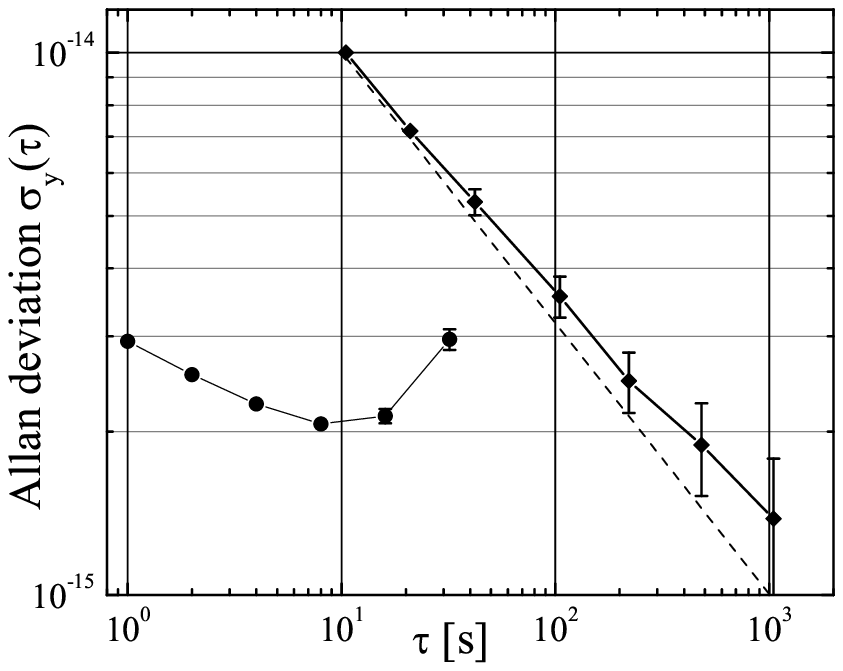}
	\caption{\label{FigStability} Circles : Fractionnal frequency instability vs integration time (characterized by the Allan standard deviation) of the microwave signal generated by the fiber-based system against the CSO at 11.932 GHz. Squares: Fiber-based femtosecond system locked onto the FO2 atomic signal, compared against the CSO (quadratic drift removed). The latter instability scales as $3.5\times10^{-14}\tau^{-1/2}$ (dashed line).}
\end{figure}

In a first experiment, the harmonic of the FOFC's repetition rate near 9.2\,GHz was sent to a nearby laboratory about 30 meters away. There, it was compared to a 9.2\,GHz microwave signal generated by a TSOFC (repetition rate $\simeq 770$\,MHz) locked onto a separate ultra-stable laser operating at 1062.5\,nm \cite{Millo2008Cavity}. The TSOFC uses a similar locking technique as the FOFC, although with a higher bandwidth of about 400\,kHz. Figure \ref{FigPhaseNoise} shows the result of a phase noise measurement of the FOFC-TSOFC beatnote signal. The Allan deviation was also measured with a bandwidth of 400\,Hz (but not represented on figure \ref{FigStability} for clarity).

In a second experiment, the FOFC's repetition rate harmonic near 11.932\,GHz \footnote{A motorized translation stage allows coarse adjustment of the repetition rate to the desired value.} was sent to the CSO/FO2 fountain laboratory 300 meters away, and there compared to the 11.932\,GHz CSO signal. In this case, the Allan deviation of the beatnote signal was measured with 10\,Hz bandwidth (see figure \ref{FigStability}). Both comparisons give a phase noise power spectral density of approximatively $10^{-9}/f^2$ [rad$^2$/Hz] (at 9.2\,GHz) for Fourier frequencies $f$ in the 0.1\,Hz-10\,Hz range and an Allan deviation about $3\times10^{-15}$ at 1\,s integration time (see figures and \ref{FigPhaseNoise} \ref{FigStability}). These performances, among the very best for microwave sources, along with the reliability and robustness of the fiber-based system qualifies it as an excellent microwave source for long-term operation of atomic fountain clocks.

In a third experiment, the low phase noise signal at 11.932\,GHz generated by our FOFC was used to replace the CSO microwave as interrogation oscillator for the FO2 atomic fountain, as shown in figure \ref{FigSchematic}. From 11.932\,GHz, our frequency synthesis \cite{Chambon2005} generates a tunable microwave signal shifted to 11.980\,GHz via a computer controlled DDS referred as DDS2. This signal is further used to generate a low phase noise 9.2\,GHz microwave signal for Ramsey spectroscopy of the cold Cesium atoms. The sequential operation of the fountain produces frequency corrections every 1.5\,s, which are applied to DDS2. The 11.980\,GHz signal is thereby locked with a bandwidth of $\simeq0.2$\,Hz onto the atomic frequency reference. The resulting primary standard referenced signal at 11.980\,GHz was compared to the CSO signal at 11.932\,GHz. The 48\,MHz difference was bridged by a low noise synthesizer. The comparison yields a fountain's instability of $3.5\times10^{-14}\tau^{-1/2}$ for integration time $\tau$ between 10\,s and 100\,s (see figure \ref{FigStability}, 10\,Hz measurement bandwidth). For integration time longer than 100\,s, the instability is limited by the flicker floor of 1-2$\times10^{-15}$ of the CSO. The short-term instability is identical to the one obtained when using the CSO as local oscillator for the FO2 fountain under the same operating conditions. The fountain instability is limited by QPN with the number of atoms of $\simeq 1\times10^{6}$ per shot (1.5\,s cycle time). The Dick effect calculated \cite{Santarelli1998} based on the phase noise shown in figure \ref{FigPhaseNoise} is below $5\times10^{-15}\tau^{-1/2}$ for the FO2 fountain current operational parameters.
 
Continuously operated fiber femtosecond optical frequency comb stabilized onto an ultra-stable laser will replace CSO as a flywheel in the near future, removing the use of cryogenics and providing an ultra stable reference in both optical and microwave domains. Furthermore, cross comparisons between fiber-based frequency comb, Ti:Sapphire-based frequency comb, and CSO microwave generation will allow full characterization and optimization of the three systems and will pave the way to extreme low-noise microwave systems for applications in radar, deep space navigation and VLBI.

\begin{acknowledgments}
Note: Upon completion of this letter, we have become aware of comparable results from S. Weyers {\em et al.} with a fountain instability of $7.4\times10^{-14}$ at 1\,s \cite{PTB}.
\end{acknowledgments}

\bibliographystyle{apsrev}

\begin{thebibliography}{}

\bibitem{WynandsMetrologia2005}
R. Wynands and S. Weyers {\em Metrologia} {\bf 42}, 64 (2005).

\bibitem{FountainItalians}
F. Levi {\em et al.}, {\em Metrologia} {\bf 43}, 545 (2006).

\bibitem{FountainGB}
K. Szymaniec {\em et al.}, {\em Metrologia} {\bf 42}, 49 (2005).

\bibitem{FountainUS}
T.P. Heavner {\em et al.}, {\em Metrologia} {\bf 42}, 411 (2005).

\bibitem{FountainSYRTE}
S. Bize {\em et al.}, {\em J. Phys. B: At. Mol. Opt. Phys.} {\bf 38}, S449 (2005).

\bibitem{OldAlphaDot}
H. Marion {\em et al.}, {\em Phys. Rev. Lett.} {\bf 90}, 150801 (2003).

\bibitem{AlphaDot}
T.M. Fortier {\em et al.}, {\em Phys. Rev. Lett.} {\bf 98}, 070801 (2007).

\bibitem{LorentzInvariance} 
N. Ashby {\em et al.}, {\em Phys. Rev. Lett.} {\bf 98}, 070802 (2007). See also Wolf et al. {\em Phys. Rev. Lett.} {\bf 96}, 060801 (2006).

\bibitem{Santarelli1999}
G. Santarelli {\em et al.}, {\em Phys. Rev. Lett.} {\bf 82}, 4619 (1999). 

\bibitem{Dick}
G. Dick, {\em in Proceedings of the Precise Time and Time
Interval Meeting}, pp. 133–147 (1988). 

\bibitem{Santarelli1998}
Santarelli {\em et al.} {\em IEEE Trans. UFFC} {\bf 45}, 887 (1998).

\bibitem{DeepSpace}
J.D. Prestage {\em et al.}, Proc. of the SPIE 667306 (2007)

\bibitem{VLBI}
S.S. Doeleman {\em et al.}, to appear in {\em Proc. of the 7th Symposium on Frequency Standard Metrology} (2008). See also S.S. Doelman {\em et al.} {\em Nature} {\bf 455}, 78 (2008). 

\bibitem{Tobar}
J.G. Hartnett {\em et al.}, {\em App. Phys. Lett.} {\bf 89}, 203513 (2006). See also C.R. Locke {\em et al.}, {\em Rev. Sci. Instrum.} {\bf 79}, 051301 (2008).

\bibitem{CavityYe2007}
A.D. Ludlow {\em et al.}, {\em Opt. Lett.} {\bf 32}, 641 (2007).

\bibitem{CavityWebster2008}
S.A. Webster {\em et al.}, {\em Phys. Rev. A} {\bf 77}, 033847 (2008).

\bibitem{Bartels2005}
A. Bartels {\em et al.}, {\em Opt. Lett.} {\bf 30}, 667 (2005).

\bibitem{PTB_Low_Noise}
B. Lipphardt {\em et al.}, {\em arXiv:0809.2150v1} (2008).

\bibitem{Chambon2005}
D. Chambon {\em et al.}, {\em Rev Sci. Inst.} {\bf 76}, 094704 (2005).

\bibitem{Millo2008Cavity}
J. Millo { \em et al.}, {\em Proc. of the 2008 IEEE Int. Freq. Cont. Symp.}, pp 110-114 (2008).

\bibitem{Jiang2008}
H. Jiang {\em et al.}, {\em JOSA B} {\bf 25}, 2029 (2008)

\bibitem{Newbury_f_2f}
B.R. Washburn {\em et al.}, {\em Opt. Lett.} {\bf 29}, Vol 3, p 250 (2004). 

\bibitem{OpticalLink}
O. Lopez {\em et al.}, {\em Eur. Phys. J. D} {\bf 48}, 3541 (2008).

\bibitem{PTB}
S. Weyers {\em et al.}, {\em arXiv:0901.2788v1}.

\end{thebibliography}

\end{document}